\documentstyle[aps,prl,preprint]{revtex}
\begin{document}
\draft
\tolerance 10000
\title {
Superconductivity of  Quasi-One-Dimensional Electrons
in Strong Magnetic Field }
\author {Yasumasa Hasegawa  and
 Mitake Miyazaki}
\address{
Faculty of Science, \\
Himeji Institute of Technology, \\
Kamigouri-chou, Akou-gun, Hyogo 678-12, Japan}

\date {\today}

\maketitle


\begin{abstract}
The superconductivity of quasi-one-dimensional electrons
in the magnetic field is studied.
The system is described as the one-dimensional electrons with no
frustration due to the magnetic field.
The interaction is assumed to be attractive between electrons in the
nearest chains, which corresponds to the lines of nodes
of the energy gap in the absence
of the magnetic field.
The effective interaction depends
on the magnetic field and the transverse momentum.
As the magnetic field becomes strong,
the transition temperature of the spin-triplet superconductivity
oscillates, while that of the
spin-singlet increases monotonically.
\end{abstract}
\noindent
\begin{center}
[ keywords:
superconductivity,
quasi-one-dimension,
organic conductors ]
\end{center}

\noindent
running head: Superconductivity in strong magnetic field

\pacs{}

\bigskip

\section{Introduction}
The reentrance of the superconductivity in
quasi-one-dimensional electrons in strong magnetic
field attracts theoretical
interest \cite{Lebed86,Burlachkov87,Dupuis93,Dupuis94,Dupuis94b}.
In the quasi-one-dimensional systems the Fermi surface is open and
Landau level quantization does not exist.
Lebed'  \cite{Lebed86}
has shown that the superconducting transition temperature reaches
to that in the absence of the magnetic field as the
magnetic field is increased. Dupuis et al.
\cite{Dupuis93,Dupuis94,Dupuis94b} have shown that
there exists the cascade of the
superconducting phases separated by first order transitions.
 The reentrance of the superconductivity is understood as follows.
When the strong magnetic field is applied along the $b$ axis,
the semi-classical orbits of electrons are localized in the
$a$-$b$ plane, where $a$ is the most conducting axis, i.e. the motion
along the $c$ axis is bounded.
Since the orbital frustration due to the magnetic field comes from the
motion of electrons in the plane perpendicular to the magnetic field,
 electrons  can make Cooper pairs without affected by the
orbital frustration and the transition temperature increases.

The reentrance of the superconductivity
in two-dimensional electrons in  strong magnetic field also
attracts theoretical interest \cite{Rasolt92,Akera91}.
This phenomena is similar to that in quasi-one-dimensional systems,
but the origin for the
reentrance of the superconductivity in two-dimension
 is different from that in quasi-one-dimension.
In two-dimensional case Cooper pairs in strong magnetic field
are formed between electrons at
the lowest Landau level,
whereas
superconductivity is stabilized in magnetic field without Landau level
quantization in quasi-one-dimensional system.

Quasi-one-dimensional electrons are realized in organic conductors,
(TMTSF)$_2$X, where anion X is PF$_6$, ClO$_4$ etc. The system can be
described as the anisotropic tight-binding model with the hopping matrix
elements, $t_a  \simeq 0.25$eV, $t_b / t_a \simeq 0.1$ and $t_c / t_a
\simeq 0.01$.
Due to the smallness of $t_c$, the magnetic field necessary to the
reentrance of the superconductivity is estimated to be
about 20T \cite{Lebed86,Dupuis93}, which can be realized in experiments.

The superconductivity in the quasi-one-dimensional
organics, (TMTSF)$_2$X, is thought to be
not the conventional BCS type even in the absence of
magnetic field.
When the nesting condition
of the Fermi surface is
satisfied sufficiently,
which is thought to be the case at low pressure,
the quasi-one-dimensional organic conductors are
not superconductors  but
insulators due to spin density wave (SDW).
Since SDW is caused by the on-site Coulomb
repulsion, the on-site repulsion is expected to be large
in this system. As a result the conventional
s-wave spin-singlet superconductivity is
not likely to be stabilized. Indeed,
the NMR relaxation rate, $T_1^{-1}$ of the
(TMTSF)$_2$ClO$_4$ shows little
enhancement just below the
transition temperature and it decreases as $T^3$ as temperature
becomes low, which strongly suggests the line nodes of the energy gap in
the superconducting state \cite{Takigawa87,Hasegawa87}.
The line nodes in the superconducting states are caused by the
attractive interaction between electrons which are not on the same site.
When the magnetic field is applied along the $y$ direction,
the $k_y$ dependence of the interaction is not affected by the magnetic
field.
Thus if the attractive interaction depends only on $k_y$,
the magnetic field dependence of the superconducting
transition temperature is the same as that in the case of the on-site
attraction. However, the $k_z$ dependence of the interaction makes the
different field dependence.
In the quasi-one dimensional case the $k_x$ dependence of the
interaction is not very
important because Fermi surface consists of two planes, $k_x \approx \pm
k_F$.
Therefore, in this paper we assume the attractive interaction
between electrons in the nearest chains in $z$ direction,
 which results in the line
node of the energy gap in the superconducting state.

In the previous paper one of the authors has  shown that
if the magnetic field is applied, the quasi-one-dimensional electrons in
the perpendicular plane
is described as the one-dimensional system and the effective
interaction depends on the magnetic field and the transverse component
of the wave vector \cite{Hasegawa95}.
When the magnetic field is applied along the $c$ axis, the
perfect nesting is recovered as long as we neglect the small
imperfectness of the order of $t_c^2/t_a$.
In that paper the on-site repulsive interaction is assumed
and the field-induced spin density wave (FISDW) is studied.
The quantum Hall effect in FISDW is understood as a result of
the $k_y$-dependent phase of the effective interaction.

  In this paper we apply the same procedure to the case of
the coexistence of the on-site repulsion and the nearest-site
attraction.
When the
 magnetic field is applied in the $y$ direction,
the imperfectness of the nesting of the Fermi surface is not changed
 and FISDW is not stabilized.
In that case the superconductivity is caused  by the attractive
part of the interaction.

\section{quasi-one-dimensional electrons in magnetic field}
We study the
 quasi-one-dimensional electrons in the magnetic field.
For simplicity, we take  $a$, $b$, and $c$ axes to be
perpendicular each other and to be along the
 $x$, $y$ and $z$ directions, respectively.
The generalization to the non-orthogonal lattice can be performed as in
the FISDW case \cite{Hasegawa95a}.
 The Hamiltonian
is written as
\begin{eqnarray}
 {\cal H}  &=& {\cal H}_0 + {\cal H}_U
\nonumber \\
 {\cal H}_0 &=&
-t_a \sum_{(i,j)_a,\sigma} e^{i\theta_{ij}}
c_{i,\sigma}^{\dagger}
           c_{j,\sigma}
\nonumber \\ & &
       - t_b \sum_{(i,j)_b,\sigma} e^{i\theta_{ij}}
      c_{i,\sigma}^{\dagger}
         c_{j,\sigma}
\nonumber \\
     & &  - t_c \sum_{(i,j)_c,\sigma} e^{i\theta_{ij}}
      c_{i,\sigma}^{\dagger}
         c_{j,\sigma}
\nonumber \\
{\cal H}_U &=&
        \sum_{< i,j>, \sigma , \sigma'}
   U_{ij} c_{i,\sigma}^{\dagger} c_{i,\sigma  }
                     c_{j,\sigma'}^{\dagger} c_{j,\sigma'}
\label{Hamiltonian}
\end{eqnarray}
where $c_{i,\sigma}^{\dagger}$ and $c_{i,\sigma}$ are
creation and annihilation operators of electrons,
$t_a$, $t_b$ and $t_c$
are the hopping matrix elements along the $a$,  $b$ and $c$ axes,
respectively, $U_{ij}$ is the interaction between electrons at $i$ and
$j$ sites and
\begin{equation}
\theta_{ij} = {2 \pi \over \phi_0 } \int_i^j {\bf A} d{\bf l}.
\end{equation}
In the above ${\bf A}$ is the vector potential and $\phi_0 = h c_0 / e$
is the flux quantum, where $c_0$ is the light velocity.
We consider the anisotropic system with the hopping matrix elements $t_a
\gg t_b \gg t_c$, which is the case in (TMTSF)$_2$X.
The nesting of the Fermi surface is not perfect due to the $t_b$ term.

In eq.(\ref{Hamiltonian}) we have neglected the Zeeman term for
simplicity.
The Zeeman term does not play any
important role for the equal-spin-pairing case of the spin triplet.
If the Zeeman energy is taken into account, the transition temperature
of the spin singlet is reduced but the superconductivity is not
completely destroyed, since half of
 the density of states is available to
make Cooper pairs for the Larkin-Ovchinikov-Frude-Ferrell
state \cite{Dupuis93,Dupuis94}.
Therefore,
the effect of the Zeeman energy for the spin singlet can be
taken into account by putting the density of states to be half
in the strong magnetic field.

In this paper the magnetic field $H$ is applied in the $y$ direction and
   the vector potential ${\bf A}$  is taken as
$
 {\bf A} = (0,\ 0,\ -  H x ).
$
 Then the non-interacting Hamiltonian is written as
\begin{eqnarray}
{\cal H}_0  &=&
\int_{k_F - {G \over 2  }}^{ k_F + { G \over 2  }}
        {d k_x \over G}
\int_{- {\pi \over b}}^{{\pi \over b}}
       { d k_y \over 2 \pi / b }
\int_{- {\pi \over c}}^{{\pi \over c}}
       { d k_z \over 2 \pi / c }
\sum_{ \sigma}
\nonumber \\
&&
 c_{\sigma}^{\dagger}({\bf k})
\left(
    \begin{array}{ccccc}
\ddots & \ddots & & & T^*          \\
\ddots & M_{-1}  &  T   & 0      & \\
       &   T^*   & M_0  & T      & \\
       & 0       &  T^* & M_{1}  & \\
 T     &         &      &        &\ddots
  \end{array}
 \right)
 c_{\sigma}({\bf k}) ,
\nonumber \\
& &
\label{hamiltonian0}
\end{eqnarray}
where
\begin{eqnarray}
M_n &=& -2 t_a \{ \cos [ a (k_x + n G ) ] - \cos (a k_F) \}
        -2 t_b \cos (b k_y) ,  \\
 T &=& - t_c \exp ({ i c k_z}),
\\
c^{\dagger}_{\sigma}
&=&
\left(  \ldots , c^{\dagger}_{\sigma} ({\bf k} - {\bf G}),
c^{\dagger}_{\sigma}  ( {\bf k}),
c^{\dagger}_{\sigma}  ( {\bf k} + {\bf G})\ldots
\right),
\end{eqnarray}
and ${\bf G} = (G, 0, 0)=(ecH/(\hbar c_0), 0, 0)$.

The creation operators of electrons can be written in terms of the
creation operators of the eigenstates of the non-interacting Hamiltonian
 in the presence of the magnetic
field.  We consider the case that the system has the open Fermi
surface and the magnetic field is not very strong in the sense that
the magnetic flux
per plaquette is much smaller than the flux quantum. Then we get $G \ll
k_F$ and
 we can treat the
electrons with $k_x ( \approx k_F )$ and $ - k_x$
independently in
${\cal H}_0$  as linear combinations
 of the eigenstates $\Psi^{\dagger}_r (n,
{\bf k} )$ and $\Psi^{\dagger}_l (n, {\bf k} )$ as \cite{Hasegawa95}
\begin{eqnarray}
c^{\dagger} ({\bf k} + m {\bf G})
&=& e^{i m c k_z} \sum_n \varphi_r (m,n)
\Psi^{\dagger}_r (n, {\bf k} )
\nonumber \\
c^{\dagger} ( - {\bf k} + m {\bf G})
&=& e^{ - i m c k_z} \sum_n \varphi_l (m,n)
\Psi^{\dagger}_l (n, - {\bf k} ),
\end{eqnarray}
where $m$ and $n$ are integers.

In this paper we take the interaction as
\begin{eqnarray}
U_{ij} = \left\{ \begin{array}{ll}
            U_0 & \mbox{if ${\bf r}_i = {\bf r}_j$ } \\
            U_1 & \mbox{if ${\bf r}_i = {\bf r}_j \pm c \hat{\bf z}$} \\
            0   & \mbox{otherwise}
                 \end{array}
          \right.
\end{eqnarray}
The Fourier-transform of the interaction is obtained as
\begin{equation}
U({\bf k}, {\bf k}') =  U_0
+ 2 U_1  \cos [ c (k_z - k_z') ].
\label{def_U}
\end{equation}
The interaction Hamiltonian is written as
\begin{eqnarray}
{\cal H}_U &=& \sum_{{\bf k},{\bf k}',{\bf q}}
               \sum_{ m, m', N}
       \sum_{\sigma, \sigma'}
 \nonumber \\
& &
 \Bigl[ U( k_z , k_z'-q_z)
c^{\dagger}_{\sigma } ({\bf k} + m {\bf G})
c^{\dagger}_{\sigma'} (-{\bf k} +{\bf q}- m {\bf G}  + N {\bf G})
 \nonumber \\
& &
\quad \times
c_{\sigma'} ( {\bf k}' + m' {\bf G} )
c_{\sigma } (-{\bf k}' +{\bf q}- m' {\bf G} + N {\bf G})
 \nonumber \\
& &
 + U( k_z , k_z')
c^{\dagger}_{\sigma } ({\bf k} + m {\bf G})
c^{\dagger}_{\sigma'} (-{\bf k} +{\bf q}- m {\bf G}  + N {\bf G})
 \nonumber \\
& &
\quad \times
c_{\sigma'} (-{\bf k}' +{\bf q}- m' {\bf G} + N {\bf G})
c_{\sigma } ( {\bf k}' + m' {\bf G} )
 \Bigr] ,
 \nonumber \\
\end{eqnarray}
where $m$, $ m' $ and $N$ are integers and
 $k_F - G/2 \leq k_x < k_F + G/2$.
The first and the second terms are the generalization of the
$g_1$ and $g_2$ terms in the
$g$-ology of the one-dimensional system \cite{Solyom79,Fukuyama85}
to the quasi-one-dimensional system.

In the previous paper \cite{Hasegawa95}
 particle-hole channel is taken into account,
since these terms make the instability of the Fermi surface into
the spin-density-wave state if the magnetic field is applied along the
$z$ direction.
 Here we consider the particle-particle
channel, which corresponds to the instability toward the
superconductivity in the case of the attractive interaction.

With the eigenstates $\Psi_{r}(n,{\bf k})$ and $\Psi_{l}(n,{\bf k})$
  the interaction
Hamiltonian is written as
\begin{eqnarray}
 {\cal H}_U &=&
 \sum_{{\bf k}, {\bf k}',{\bf q}}
 \sum_{n_1, n_2, n_3, n_4}
 \sum_{\sigma_1, \sigma_2, \sigma_3, \sigma_4}
U_{\sigma_1, \sigma_2, \sigma_3, \sigma_4}
 (n_1, n_2, n_3, n_4, k_z, k_z',q_z)
\nonumber \\
& & \times
\Psi_{r \sigma_1}^{\dagger}(n_1,{\bf k})
\Psi_{l \sigma_2}^{\dagger}(n_2,- {\bf k} +{\bf q})
\nonumber \\
& &
 \times
\Psi_{l \sigma_3}(n_3,- {\bf k}'+{\bf q})
\Psi_{r \sigma_4}(n_4,{\bf k}')
\end{eqnarray}
where
\begin{eqnarray}
&U&_{\sigma_1, \sigma_2, \sigma_3, \sigma_4}
 (n_1, n_2, n_3, n_4, k_z, k_z',q_z)
\nonumber \\
 & &= \sum_{m, m', N}
\exp \left[ i \left\{ 2 m c ( k_z - {q_z \over 2})
                    - 2 m'c ( k_z'- {q_z \over 2})
 -N  c (k_z -k_z') \right\} \right]
\nonumber \\
& &
\times
\varphi_r(m,n_1) \varphi_l(-m+N,n_2)
\nonumber \\
& &
\times
 \varphi_l^*(-m'+N,n_3)
\varphi_r^*(m',n_4)
\nonumber \\
& &
\times
\Bigl[  - U(k_z, -k_z' + q_z )
\delta_{\sigma_1, \sigma_3}
\delta_{\sigma_2, \sigma_4}
+ U(k_z,k_z')
\delta_{\sigma_1, \sigma_4}
\delta_{\sigma_2, \sigma_3}
\Bigr]
\nonumber \\
\label{hint}
\end{eqnarray}

The coefficients $\varphi_r (m, n)$ and $\varphi_l (m, n)$ can be
calculated by diagonalizing the matrix in eq.(\ref{hamiltonian0})
numerically as we have done in the previous paper \cite{Hasegawa95}.
 In this paper, instead of diagonalizing the matrix numerically,
 we use the approximation that
the dispersion in $k_x$ is taken  to be linear, i.e.,
\begin{equation}
M_n \approx v_F (k_x + n G - k_F) - 2 t_b \cos (b k_y),
\end{equation}
where  $v_F = 2 t_a a \sin a k_F$ is the Fermi velocity.
With this approximation the eigenvalues are obtained as
\begin{equation}
\xi(n, k_x, k_y) = v_F (k_x + nG - k_F) - 2 t_b \cos ( b k_y )
\end{equation}
and the eigenstates are given with the coefficients
\begin{eqnarray}
\varphi_r(m,n) &=& J_{-n+m}\left( z  \right)
\end{eqnarray}
and
\begin{eqnarray}
\varphi_l(m,n) &=& J_{ n-m}\left( z \right),
\end{eqnarray}
where $J_n(z)$ is the Bessel function
and $z = 2 t_c / (v_F G)$.
We can perform the $m$ and $m'$ summation in eq.(\ref{hint})
by using the identity,
\begin{eqnarray}
\sum_{m= - \infty}^{\infty }
e^{ i m Q} J_{m-N}(z) J_m(z) = e^{  i {N \over 2} (Q + \pi)}
J_N(2z \sin {Q \over 2} ) ,
\end{eqnarray}
and get
\begin{eqnarray}
U_{\sigma_1, \sigma_2, \sigma_3, \sigma_4}
 (n_1, n_2, n_3, n_4, k_z, k_z',q_z)
 &=& \sum_{ N}
 \exp \left[ i \left\{ (n_1-n_2) c (k_z-{q_z \over 2})
\right. \right. \nonumber \\
& &
 + (n_3-n_4) c (k_z'-{q_z \over 2})
\nonumber \\
& & \left. \left.
 - (n_1+n_2+n_3+n_4-2N)
{\pi \over 2} \right\} \right]
\nonumber \\
& &
 \times
    J_{N-n_1-n_2} ( 2z \sin c (k_z-{q_z \over 2} ) )
\nonumber \\
& &
 \times
    J_{N-n_3-n_4} (-2z \sin c (k_z'-{q_z \over 2} ) )
\nonumber \\
& &
\times
\Bigl[  - U(k_z, -k_z' + q_z )
\delta_{\sigma_1, \sigma_3}
\delta_{\sigma_2, \sigma_4}
+ U(k_z,k_z')
\delta_{\sigma_1, \sigma_4}
\delta_{\sigma_2, \sigma_3}
\Bigr] .
\nonumber \\
\
\label{Usigma}
\end{eqnarray}
In the strong magnetic
field the instability of the Fermi surface
toward forming Cooper pairs
occurs due to terms with
$q_x = 0$, $n_1 = -n_2$ ($ = n$) and $n_4 = -n_3$ ($ = n'$).
Taking only these terms corresponds to the quantum limit
approximation \cite{Dupuis94b}.
 In this paper we take this approximation,
which is justified when $T \ll v_F G$.

The order parameter is defined as \cite{Sigrist91}
\begin{eqnarray}
\Delta_{\sigma_2, \sigma_1} (n,k_z,q_z)
&=&
 - \sum_{n', {\bf k}'} \sum_{\sigma_1, \sigma_2,
\sigma_3, \sigma_4}
U_{\sigma_1, \sigma_2,\sigma_3, \sigma_4}
 (n, -n, -n', n', k_z, k_z', q_z )
\nonumber \\
& & \times
< \Psi_{l \sigma_3 } (-n', -{\bf k}' + {\bf q} )
  \Psi_{r \sigma_4 } ( n',  {\bf k}'  ) > .
\end{eqnarray}
The mean-field Hamiltonian is written as
\begin{eqnarray}
{\cal H}_{MF} &=& \sum_{n,{\bf k},\sigma}
\Bigl[ \xi (n,k_x, k_y)
\{ \Psi_{r \sigma }^{\dagger} (n,{\bf k})
   \Psi_{r \sigma } (n,{\bf k})
\nonumber \\
& &
+ \Psi_{l \sigma}^{\dagger} (-n, - {\bf k})
  \Psi_{l \sigma} (-n, - {\bf k})
\} \Bigr]
\nonumber \\
& &
- \sum_{n, {\bf k}, {\bf q}, \sigma_1, \sigma_2}
\Bigl[ \Delta_{\sigma_2 \sigma_1 } (n, k_z, q_z)
\Psi_{r \sigma_1 }^{\dagger} (n,{\bf k})
\Psi_{l \sigma_2}^{\dagger} (-n, - {\bf k} + {\bf q})
+ H.c.
  \Bigr].
\end{eqnarray}
Note that $\xi(n,k_x, k_y)$ does not depend on $k_z$, which means
the one-dimensional motion in the $x$-$z$ plane in the semi-classical
picture.
\section{transition temperature}
The transition temperature is given by the linearized gap equation,
\begin{eqnarray}
\Delta_{\sigma_2 \sigma_1 } (n, k_z,q_z)
&=&
- \sum_{n'}
\int_{k_F - { G \over 2}}^{k_F + {G \over 2}}  {d k_x' \over G}
\int_{- {\pi \over b} }^{ \pi \over b} {d k_y' \over 2 \pi /b}
\int_{-{ \pi \over c} }^{ \pi \over c } {d k_z' \over 2 \pi /c}
\sum_{\sigma_3, \sigma_4}
\nonumber \\
& &
U_{\sigma_1, \sigma_2,\sigma_3, \sigma_4}
 (n, -n, -n', n', k_z, k_z', q_z )
{\tanh {\xi(n',k_x', k_y') \over 2T} \over 2 \xi(n',k_x',k_y')}
\Delta_{\sigma_3 \sigma_4 } (n', k_z',q_z) .
\label{gapeq0}
\end{eqnarray}
This equation is simplified by defining $\Delta_{m,\sigma_1 \sigma_2}$ by
\begin{eqnarray}
\Delta_{\sigma_2 \sigma_1}(n,k_z,q_z)
&=&
e^{2 i n c (k_z - {q_z \over 2} ) } \sum_{m = - \infty} ^{\infty}
 e^{i m c (k_z - {q_z \over 2} ) } \Delta_{m, \sigma_2 \sigma_1} ,
\end{eqnarray}
and
using
\begin{eqnarray}
\sum_{n'}
\int_{k_F - { G \over 2}}^{k_F + {G \over 2}}  {d k_x' \over G}
\int_{- {\pi \over b} }^{ \pi \over b} {d k_y' \over 2 \pi /b}
{\tanh{\xi (n',k_x', k_y') \over 2 T} \over 2 \xi (n' ,k_x', k_y')}
 &=&
{N(0) \over 2 } \ln {2 \gamma \omega_c \over \pi T} ,
\label{sum0}
\end{eqnarray}
where
 $N(0)$ is the density of states for one spin
at the Fermi energy, $\omega_c$
is the cut-off energy, and
$\gamma$ is the exponential of the Euler constant.
The factor $1/2$ in the right hand side of eq.(\ref{sum0}) comes from
the fact that only half of the Fermi surface ($k_x \approx + k_F$) is taken
in the summation.

\subsection{spin singlet}
We can treat the spin singlet case  and the spin triplet case
separately. In this section we consider the spin singlet case.
For the spin singlet case the order parameter is defined by
\begin{eqnarray}
\Delta_m^{(s)} = {1 \over 2} ( \Delta_{m,\uparrow \downarrow}
                -\Delta_{m,\downarrow \uparrow} )
\end{eqnarray}
and the transition temperature is determined by the equation,
\begin{eqnarray}
\Delta_m^{(s)}=   N(0) \ln {1.13 \omega_c \over T}
 \sum_{m'= - \infty}^{\infty} M_{m,m'}^{(s)} \Delta_{m'}^{(s)} ,
\end{eqnarray}
where
\begin{eqnarray}
M_{m,m'}^{(s)} &=&
  - \int_{- {\pi \over c}}^{ \pi \over c} {d k_z \over 2\pi /c }
    \int_{- {\pi \over c}}^{ \pi \over c} {d k_z' \over 2\pi /c}
 e^{-i c (m k_z - m' k_z')}
\nonumber \\
& & \times
J_0(2 z \{\sin c k_z - \sin c k_z' \})
\nonumber \\
& & \times
[
U_0 + 2 U_1 \cos c k_z \cos c k_z'
]
\end{eqnarray}
If the maximum eigenvalue of the matrix $M_{m,m'}^{(s)}$, which we define
$g^{(s)}$, is positive,
the transition temperature is given by
\begin{eqnarray}
T_c^{(s)} = 1.13 \omega_c \exp ( - {1 \over N(0) g^{(s)}} ).
\end{eqnarray}
Using the identity for the Bessel function
\begin{eqnarray}
J_N(z) = \int_0^{2 \pi} {d\theta \over 2\pi}
\exp [ i (  N \theta - z \sin \theta ) ] ,
\end{eqnarray}
 we get
\begin{eqnarray}
   M_{m,m'}^{(s)} &=& - U_0 L_{m,m'} - U_1 C_{m,m'} ,
\end{eqnarray}
where
\begin{eqnarray}
L_{m,m'} = \int_0^{2 \pi} {d\theta \over 2\pi} J_m (2 z \sin \theta)
J_{m'}  (2 z \sin \theta)
\end{eqnarray}
and
\begin{eqnarray}
    C_{m,m'} =  {1 \over 2}
  ( L_{m+1,m'+1} + L_{m-1,m'-1} + L_{m+1,m'-1} + L_{m-1,m'+1} ).
\end{eqnarray}

First we consider
the case  $U_0 < 0$ and $U_1 = 0$ ( i.e. on-site attraction),
which has been studied by
Lebed' \cite{Lebed86,Burlachkov87}
and Dupuis et al. \cite{Dupuis93,Dupuis94,Dupuis94b}.
Noting that
\begin{equation}
L_{m,m'} = \sum_{N=-\infty}^{\infty} T_{m,N}^* T_{N,m'}
\end{equation}
where
\begin{eqnarray}
T_{N,m'}
&=&
\int_{-\pi}^{\pi}{d \theta' \over 2 \pi} e^{i N \theta}
  J_{m'}(2z \sin \theta ')
\nonumber \\
&=&
\int_{-{\pi \over c}}^{\pi \over c}{d k_z \over 2 \pi/c} e^{i m' c k_z}
  J_{N}(2z \sin c k_z )
\end{eqnarray}
we can show that
 all eigenvalues of the matrix $L_{m,m'}$ are positive or zero.
Since $L_{m,m'}=0$ if one of $m$ and $m'$ is odd and the other is
even, the matrix  $( M_{m,m'})$ is divided into two parts, even and odd.
When the magnetic field is changed, the maximum eigenvalue is obtained by
the even part or the odd part, resulting in the
cascade transition into the superconducting state,  as shown by Dupuis
et al. \cite{Dupuis93,Dupuis94,Dupuis94b}.
 In Fig.1 we plot
the effective coupling constant $g^{(s)}/ | U_0 |$
 as a function
of $h =  H / H_0$, where $H_0 =  2 t_c \hbar c_0 / (v_F e b c )$.
 When we take  the
parameters as $t_a \approx 3000$K, $t_c \approx 20$K,
 $a \approx 7$\AA, $c \approx 14$\AA \  and $a k _F = \pi /4$, we get
$H_0 \approx 10$T.

Next we consider the case $U_0 > 0$ and $U_1 < 0$. Since
\begin{eqnarray}
 L_{m,m'}= (-1)^{m} L_{-m,m'} = (-1)^{m'} L_{m,- m'}
\end{eqnarray}
and
\begin{eqnarray}
 C_{m,m'} = - (-1)^{m} C_{-m,m'} = - (-1)^{m'} C_{m,- m'} ,
\end{eqnarray}
we get $L C = C L = 0$, i.e.
the eigenstates of  $M$ are the
simultaneous eigenstates of $L$ and $C$ and at least
one of the eigenvalues of
$L$ and $C$ for each eigenstate is zero.
All eigenvalues of
$C_{m,m'}$ are positive or zero, which can be shown by noting
\begin{eqnarray}
 C_{m,m'} = \sum_{N=-\infty}^{\infty}
 {1 \over 2}
 (T^*_{m+1,N} + T^*_{m-1,N} ) (T_{N,m+1} + T_{N,m-1} ).
\end{eqnarray}
Thus $g^{(s)}$ is obtained by calculating the maximum eigenvalue of
$C_{m,m'}$ in the case of $U_0 > 0$ and $U_1 < 0$. In this case the
transition temperature does not depend on $U_0$.
In Fig.2 $g^{(s)} / | U_1|$ is plotted as a function of $h$.
In contrast to Fig.1 the effective coupling constant does not oscillate
as the magnetic field is increased. Only in the small region of $h
\approx 0.2$ the largest eigenvalues for the even part and the odd part
of the matrix $C_{m,m'}$  intersect.

\subsection{spin triplet}
In this section we consider the spin triplet pairing case.
The linearized gap equation for the spin triplet pairing is given as
\begin{equation}
\Delta_m^{(t)} = N(0) \ln {1.13 \omega_c \over T}
\sum_{m'= - \infty}^{\infty} M_{m,m'}^{(t)} \Delta_m^{(t)},
\end{equation}
where
$\Delta_m^{(t)}$ is either $\Delta_{m,\uparrow \uparrow}$,
$(\Delta_{m,\uparrow \downarrow} + \Delta_{m,\downarrow \uparrow}) /2$
or $\Delta_{m,\downarrow \downarrow}$ and
\begin{eqnarray}
M_{m,m'}^{(t)} &=&
 - { U_1 \over 2}
  ( L_{m+1,m'+1} + L_{m-1,m'-1} - L_{m+1,m'-1} - L_{m-1,m+1} )
\end{eqnarray}
If the maximum eigenvalue of the matrix
$M_{m,m'}^{(t)}$
is positive, we define it as $g^{(t)}$ and we get
the transition temperature as
\begin{eqnarray}
T_c^{(t)} = 1.13 \omega_c \exp (- {1 \over N(0) g^{(t)}})
\end{eqnarray}
In Fig. 3 we plot $g^{(t)} / | U_1 |$ vs. $h$.
The effective coupling constant is given by the eigenvalue
of the even part or odd part of the matrix $M_{m,m'}^{(t)}$ depending on
$h$, as in the case of the spin singlet for the on-site attractive
interaction.

\section{conclusion}
In this paper we have studied the transition to superconductivity in the
quasi-one-dimensional systems in the
magnetic field applied along the $y$ direction.
The magnetic field per plaquette is much smaller than the flux quantum
but strong enough so that we can apply the quantum limit
approximation, which is realized in quasi-one-dimensional organic
superconductors in the magnetic field of the order of $10$T.
With this condition the system is treated as one-dimension
with the effective interaction depending on the magnetic field.
We studied the case that the interaction
between electrons is repulsive for the on-site electrons and attractive
for electrons at  the nearest sites in the $z$ direction.
We obtained that the effective
coupling constant for the spin triplet oscillates as the magnetic field
is increased, while that
 for the spin singlet increases monotonically.

\acknowledgments
This work is financially supported by the Grant-in-Aid for Scientific
Research
on the priority area ``Novel Electronic States in Molecular Conductors''
from the Ministry of Education.
%


\begin{figure}
\caption{
 The effective coupling constant for the spin singlet in the case of
on-site interaction ($U_0 < 0$) as a function of the magnetic field $h =
H / H_0$. The effective interaction is given by the eigenvalues for the
even part of the matrix (solid lines) and the odd part (broken lines).
}
\label{Fig.1}
\end{figure}

\begin{figure}
\caption{
The effective coupling constant for the spin singlet in the case of
attraction interaction between nearest sites ($U_1 < 0$)
as a function of the
magnetic field. The solid lines show the largest and the second largest
eigenvalues for the even part of the matrix and the broken lines show
the eigenvalues for the odd  part.
}
\label{Fig.2}
\end{figure}

\begin{figure}
\caption{
The effective coupling constant for the spin triplet in the case of
attraction interaction between nearest sites ($U_1 < 0$)
 as a function of the
magnetic field.
The solid lines show the largest and the second largest
eigenvalues for the even part of the matrix and the broken lines show
the eigenvalues for the odd part.
}
\label{Fig.3}
\end{figure}
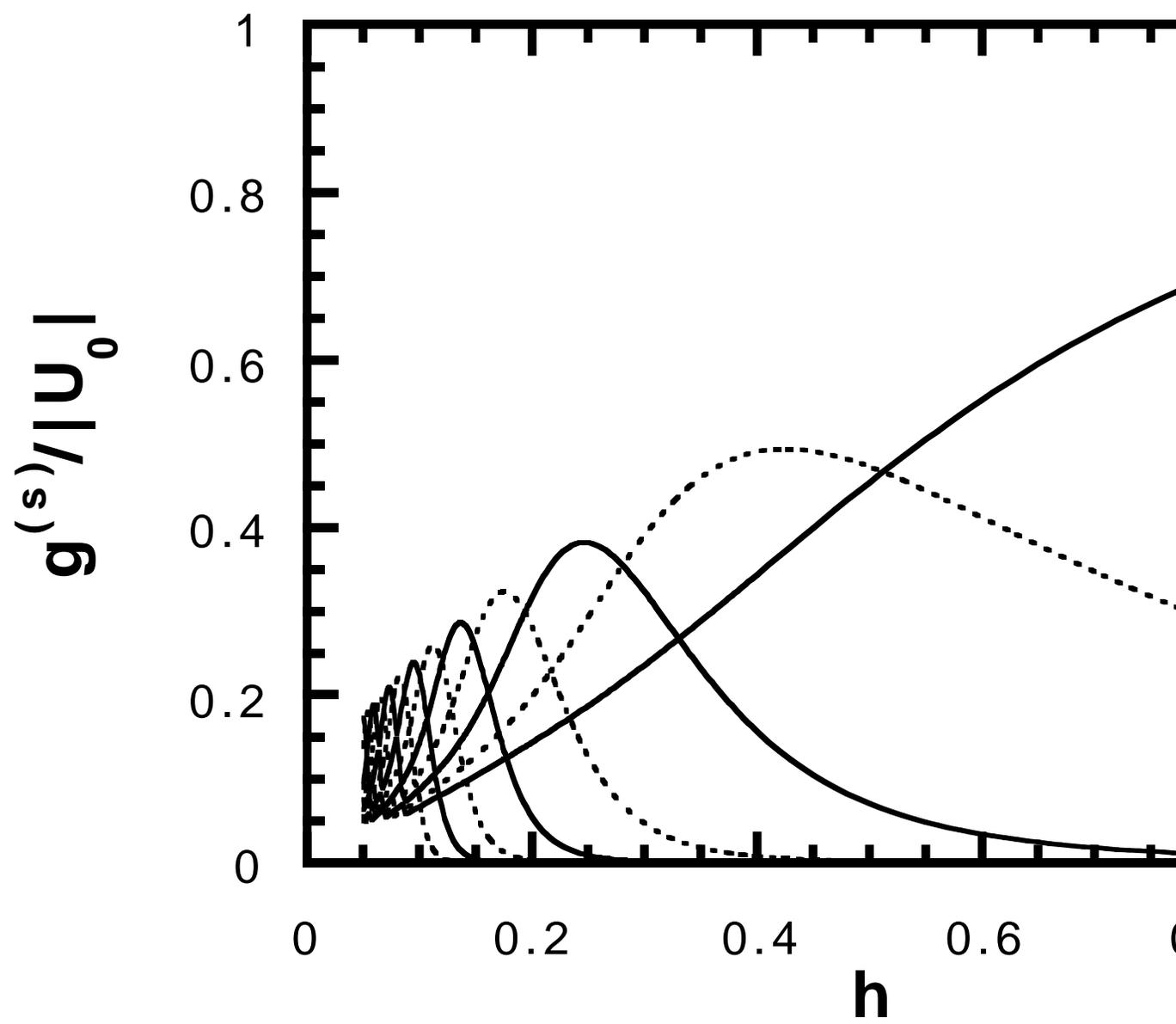
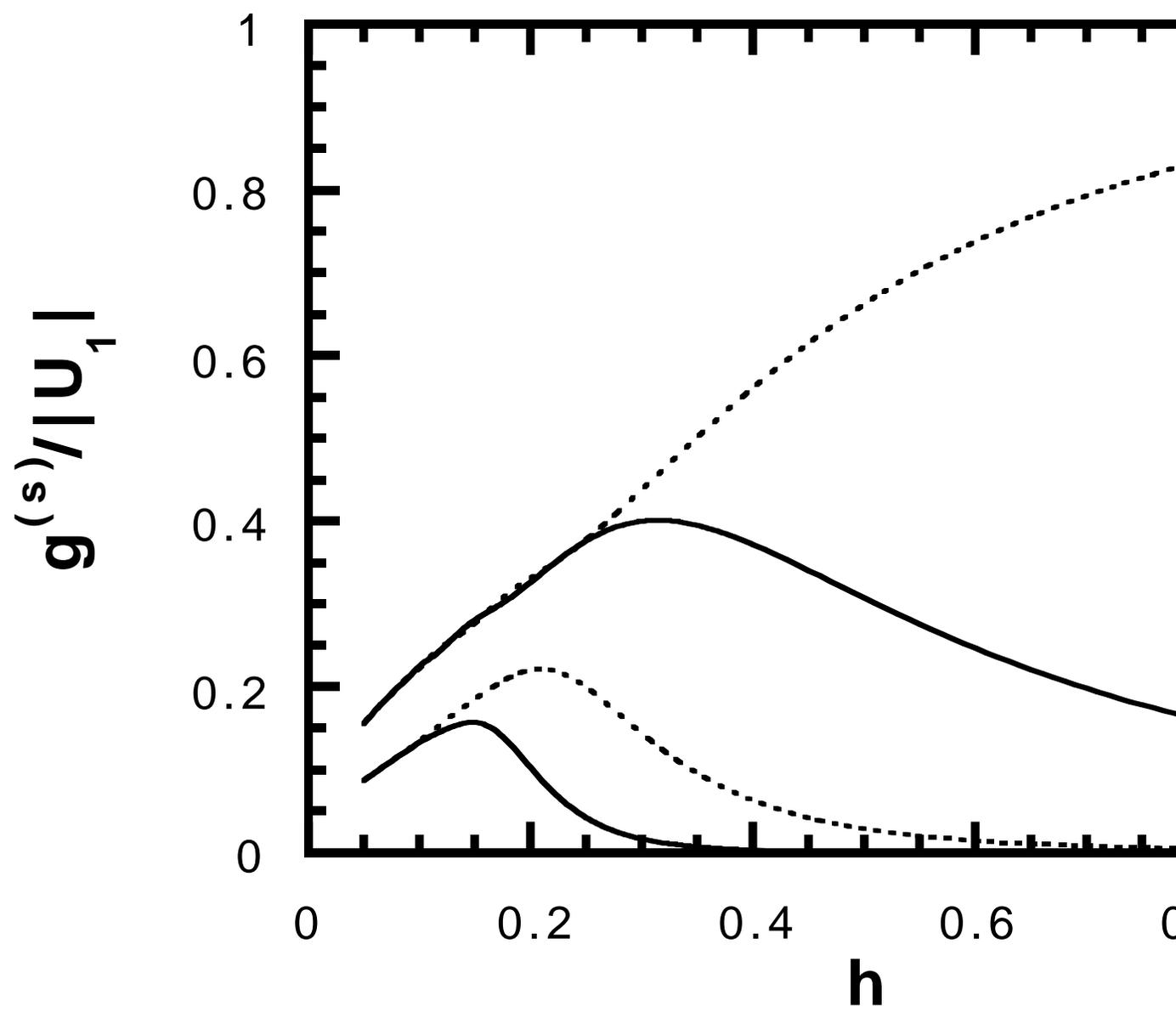
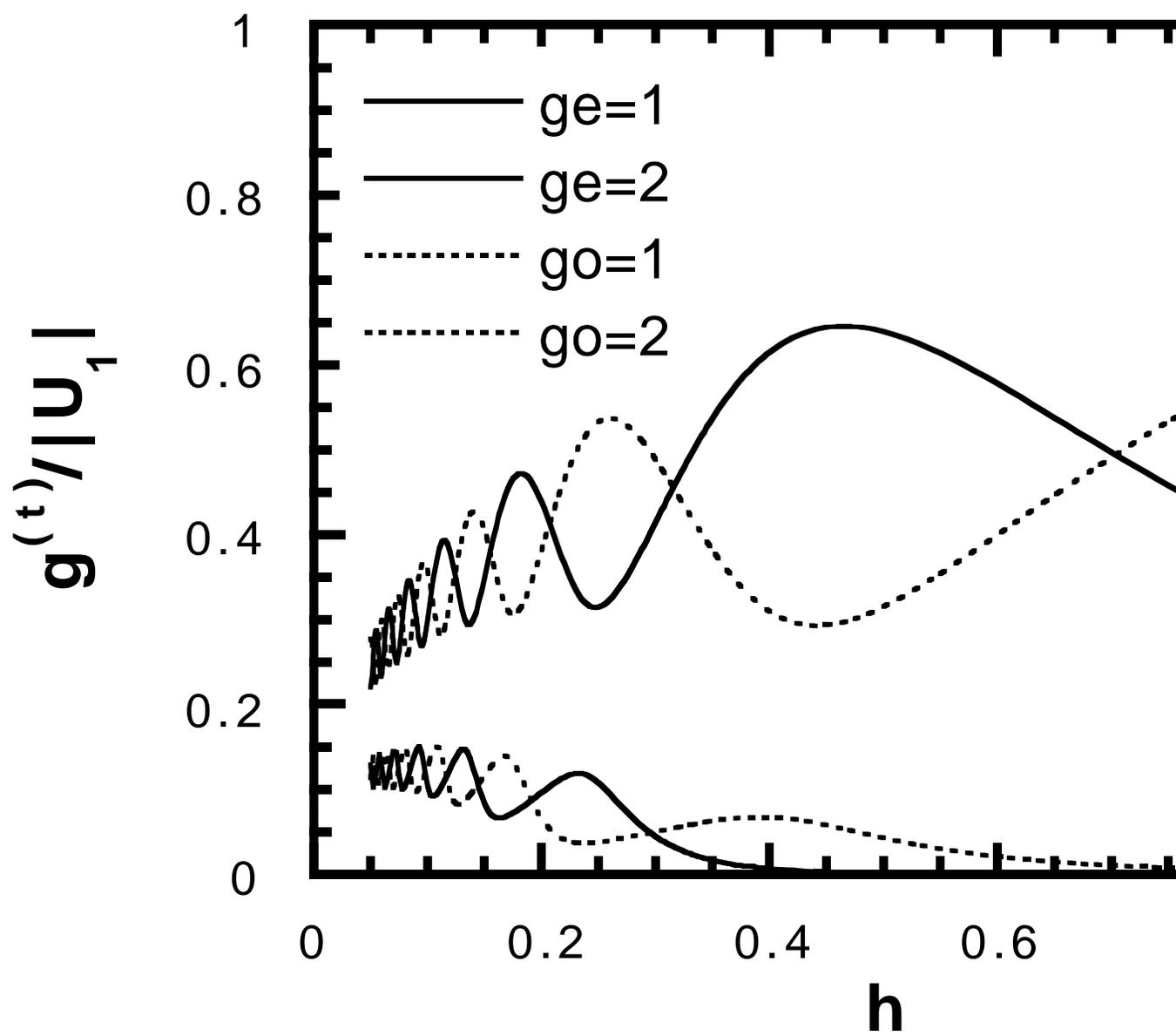
\end{document}